\documentclass{article}  

\usepackage{arxiv}              
\usepackage[utf8]{inputenc}
\usepackage[T1]{fontenc}
\usepackage{hyperref}
\usepackage{url}
\usepackage{booktabs}
\usepackage{amsfonts}
\usepackage{nicefrac}
\usepackage{microtype}
\usepackage{graphicx}
\graphicspath{ {./images/} }
\usepackage{tfrupee}            
\usepackage{amsmath}
\usepackage{float}
\usepackage{adjustbox}  
\usepackage{placeins}

\title{Quantifying Inter-Annual Seasonal Drift in Tomato Prices Using
       Dynamic Time Warping: Evidence from Kolar Market}

\author{
  Manojkumar Patil \\
  Dept.\ of Agricultural Economics, UAS, Bengaluru, India \\
  \texttt{patil.manojkumar@hotmail.com}
  \AND
  Lalith Achoth \\
  Dept.\ of Agricultural Economics, UAS, Bengaluru, India \\
  \AND
  K.\ B.\ Vedamurthy \\
  Dept.\ of Dairy Business Management, Dairy Science College, KVAFSU, Bengaluru, India \\
  \AND
  K.\ B.\ Umesh \\
  Dept.\ of Agricultural Economics, UAS, Bengaluru, India \\
  \AND
  Siddayya \\
  Institute of Agribusiness Management, UAS, Bengaluru, India \\
  \AND
  M.\ N.\ Thimme Gowda \\
  Dept.\ of Agrometeorology, UAS, Bengaluru, India
}

\begin{document}
\maketitle

\begin{abstract}
Tomato prices in Kolar market exhibit high volatility alongside recurring
seasonal patterns, but the consistency of these patterns across years remains
unclear. This study analysed weekly tomato prices and arrivals from 2010--2024
to quantify inter-annual variability using descriptive statistics, seasonal
indices, and Dynamic Time Warping (DTW). Descriptive analysis confirmed extreme
fluctuations (CV = 77\% for prices, 102\% for arrivals) with positive skewness
and heavy tails, indicating frequent extreme events. Seasonal indices revealed
recurring intra-year cycles, but year-to-year alignment varied substantially.
DTW analysis for 2021--2024 quantified pattern similarity, showing that
2022--2023 had the highest alignment (DTW distance: 23,258) despite extreme
price spikes (max: \rupee~7,429), whereas 2021--2022 exhibited the weakest
alignment (distance: 39,049), reflecting structural shifts in market dynamics.
Path length metrics indicated minimal temporal warping in 2022--2023 (71
points) versus extensive alignment in 2021--2022 (83 points). These results
demonstrate that while seasonal patterns recur, their temporal consistency is
not fixed, highlighting the need for forecasting models that adapt to both
magnitude volatility and temporal shifts. The study also illustrates the
utility of DTW for agricultural price analysis and the limitations of relying
solely on fixed seasonal patterns in volatile commodity markets.
\end{abstract}


\section{Introduction}\label{sec:intro}

Tomato is India's most important vegetable crop, yet its market is notorious
for intra-year price swings that often exceed 200 per cent
\cite{rediff2024,pib2024_tomato,thehindu2024,niti2024}. Perishability,
concentrated harvest windows and thin cold-chain capacity expose smallholders
to large income risk \cite{roy2024,mittal2007}. Forecasting models from SARIMA
to LSTM capture mean seasonality but lose accuracy during volatile spells, with
errors above 20 per cent \cite{manogna2025}. A neglected issue is whether the
calendar of seasonality itself changes. Classical seasonal indices assume
identical timing each year; if peaks slide by several weeks, fixed-dummy
forecasts fail. Dynamic Time Warping (DTW) allows elastic alignment of two
series and has been used in energy and equity research; applications to
perishable food prices remain rare.

This study examines weekly tomato prices and arrivals at Kolar APMC
(2010--2024) to (i) profile volatility, (ii) extract seasonal indices, and
(iii) quantify year-to-year calendar drift with DTW over 2021--2024. Kolar is
South India's dominant tomato supply hub, so its price discovery influences
neighbouring states.

The study contributes: (a) a DTW-based diagnostic for inter-annual pattern
consistency, (b) empirical evidence that tomato seasonal timing is recurrent
but temporally unstable, and (c) guidance for adaptive forecasting that
tolerates both magnitude and calendar shocks.


\section{Methodology}\label{sec:method}

\subsection{Study Area and Data}
This study used secondary time-series data on tomato arrivals and modal prices
from the Kolar Agricultural Produce Market Committee (APMC), Karnataka, India.
Weekly tomato arrivals (quintals) and modal prices (\rupee/quintal) for January
2010 to December 2024 (792 observations) were obtained from
Krishimaratavahini and Agmarknet. Modal price represents the value at which the
largest quantity is traded on a given day, aggregated to the week-ending date.
This measure is commonly used in agricultural market analysis, as it more
accurately reflects typical transaction prices than minimum or maximum values.

\subsection{Data Cleaning}
Missing weekly observations ($<2\%$) were interpolated using cubic splines.
Outliers, defined as values beyond
\[
  Q_1 - 3 \times \mathrm{IQR} \quad \text{or} \quad Q_3 + 3 \times \mathrm{IQR},
\]
were cross-checked against market bulletins and retained if verified as genuine
price or arrival spikes. Stationarity of the series was assessed using the
Augmented Dickey--Fuller (ADF) test, which rejected the unit-root hypothesis
for log-price differences ($p < 0.01$).

\subsection{Descriptive Statistics}
Comprehensive descriptive statistics were computed to characterise the central
tendency, dispersion, and distributional properties of weekly tomato arrivals
and modal prices. The metrics calculated included the mean, median, standard
deviation, coefficient of variation (CV), skewness, kurtosis, and the
Jarque--Bera test for normality. These measures provided insights into the
typical values, variability, asymmetry, and tail behaviour of the series,
helping to assess volatility and the prevalence of extreme events in the market.

\subsection{Seasonal Indices}
Seasonal indices were calculated to identify recurring intra-year price and
arrival patterns. The analysis used ISO week numbering (weeks 1--52/53) to
align observations across different years \cite{kendall1983}. This approach
assumes that a time series can be decomposed into trend, seasonal, cyclical,
and irregular components \cite{makridakis1998}. The seasonal indices were
computed separately for tomato arrivals and modal prices to examine whether
supply and price seasonality patterns align or diverge.

\subsection{Dynamic Time Warping (DTW)}
Dynamic Time Warping (DTW) was applied to align tomato price trajectories
across different years. DTW measures the similarity between two temporal
sequences that may vary in timing, allowing detection of interannual
irregularities and structural shifts in price patterns. Price data from each
year were aligned against a reference year for analysis. Unlike
correlation-based measures that require point-to-point temporal alignment, DTW
allows flexible matching of time series with temporal shifts, compressions, or
expansions \cite{sakoe1978,berndt1994}. This makes DTW particularly suitable
for capturing seasonal deviations and timing differences in market price
dynamics.

\subsubsection{DTW Algorithm}
Let two time series be defined as
\[
  X = \{x_1, x_2, \dots, x_n\}, \quad
  Y = \{y_1, y_2, \dots, y_m\}.
\]
The DTW algorithm finds an optimal warping path between $X$ and $Y$ that
minimises the cumulative distance between aligned elements.

\paragraph{Step 1: Distance Matrix.}
A local distance matrix $D$ of size $n \times m$ is computed as:
\[
  d(i,j) = |x_i - y_j|.
\]
For multivariate series, the Euclidean distance is used:
\[
  d(i,j) = \sqrt{\sum_{k=1}^{p} \left(x_{i,k} - y_{j,k}\right)^2}.
\]

\paragraph{Step 2: Cumulative Cost Matrix.}
A cumulative cost matrix $\gamma$ is computed recursively:
\[
  \gamma(i,j) = d(i,j) + \min
  \begin{cases}
    \gamma(i-1,j-1) & \text{(diagonal)} \\
    \gamma(i-1,j)   & \text{(vertical)} \\
    \gamma(i,j-1)   & \text{(horizontal)}
  \end{cases}
\]
with boundary conditions:
\[
  \gamma(1,1) = d(1,1), \quad
  \gamma(i,1) = \sum_{k=1}^{i} d(k,1), \quad
  \gamma(1,j) = \sum_{k=1}^{j} d(1,k).
\]

\paragraph{Step 3: Optimal Warping Path.}
The optimal warping path
\[
  W = \{w_1, w_2, \dots, w_K\}, \quad w_k = (i_k, j_k),
\]
is obtained by backtracking from $\gamma(n,m)$ to $\gamma(1,1)$, always
following the predecessor with the minimum cumulative cost.

\paragraph{Step 4: DTW Distance.}
The DTW distance between $X$ and $Y$ is:
\[
  \mathrm{DTW}(X,Y) = \frac{\gamma(n,m)}{K},
\]
where $K$ is the length of the warping path. A lower DTW distance indicates
higher similarity between the temporal patterns of the two series.

\subsubsection{Implementation Details}
Pair-wise alignments of weekly modal prices were performed for consecutive
years 2021--2024 using the Python \texttt{darts} library. A full warping window
(no Sakoe--Chiba band) and Euclidean local cost were used initially; robustness
was checked with $\pm4$-week bands. Year-pair rankings did not change, so
unrestricted results are reported. Metrics extracted: DTW distance, mean
point-wise distance, path length, and warped series.

\subsection{Software and Statistical Tools}
All analyses were performed in Python~3.10 using \texttt{pandas},
\texttt{NumPy}, \texttt{SciPy}, \texttt{statsmodels}, \texttt{darts},
\texttt{Plotly}, \texttt{matplotlib}, and \texttt{seaborn}. The significance
level was set at $\alpha = 0.05$.


\section{Results and Discussion}\label{sec:results}
\subsection{Descriptive Statistics and Distributional Analysis}
Table~\ref{tab:desc_stats} summarises the weekly tomato arrivals and modal
prices in Kolar Market from 2010 to 2025. Weekly tomato arrivals and modal
prices exhibited substantial variability, with coefficients of variation of
101.9\% for arrivals and 77.1\% for prices, far exceeding levels typically
observed in stable commodity markets \cite{reddyreddy2018,chandtewari2017}.
Positive skewness (2.02 for arrivals, 2.70 for prices) and high kurtosis
(4.38 for arrivals, 11.33 for prices), along with Jarque--Bera $p$-values
$<0.001$, indicate frequent extreme values and heavy-tailed distributions,
consistent with findings from other perishable commodity markets
\cite{ali2023,miljkovic2023,cont2001}.

These characteristics suggest that extreme price movements occur more
frequently than would be expected under normal assumptions, making classical
models such as ARIMA inadequate for capturing full market dynamics
\cite{box2015,makridakis1998}. Robust modelling approaches or appropriate
transformations, including logarithmic or Box--Cox, are recommended
\cite{boxcox1964,taylor2018}. Modern forecasting frameworks --- SARIMA,
Prophet, XGBoost, and LSTM --- have demonstrated superior performance in
capturing non-linear and volatile patterns in agricultural markets
\cite{chen2016,hochreiter1997,mishra2021,yadav2025}.

The heavy-tailed distributions also imply that prediction intervals should be
wider than those based on normality, which is crucial for forecasting and risk
management in volatile markets \cite{gilbert2010,miljkovic2023}. These insights
provide guidance for storage planning, market interventions, and price
stabilisation strategies, helping to mitigate income risks for farmers and
traders.

\begin{table}[htbp]
\centering
\caption{Descriptive statistics of tomato arrivals and prices in Kolar Market
         (2010--2025)}
\label{tab:desc_stats}
\begin{tabular}{lcc}
\toprule
\textbf{Metric} & \textbf{Arrivals (qtl)} & \textbf{Modal Price (\rupee/qtl)} \\
\midrule
Count         & 792     & 792    \\
Mean          & 6{,}772 & 1{,}140 \\
Std           & 6{,}899 & 880    \\
CV (\%)       & 101.89  & 77.13  \\
Skewness      & 2.02    & 2.70   \\
Kurtosis      & 4.38    & 11.33  \\
Min           & 361     & 153    \\
25th pctile   & 2{,}016 & 588    \\
Median        & 4{,}634 & 864    \\
75th pctile   & 8{,}569 & 1{,}424 \\
Max           & 38{,}277 & 7{,}429 \\
JB $p$-value  & $<0.001$ & $<0.001$ \\
\bottomrule
\end{tabular}
\end{table}

\subsection{Seasonal Pattern Analysis}

\begin{figure}[htbp]
\centering
\includegraphics[width=0.95\linewidth]{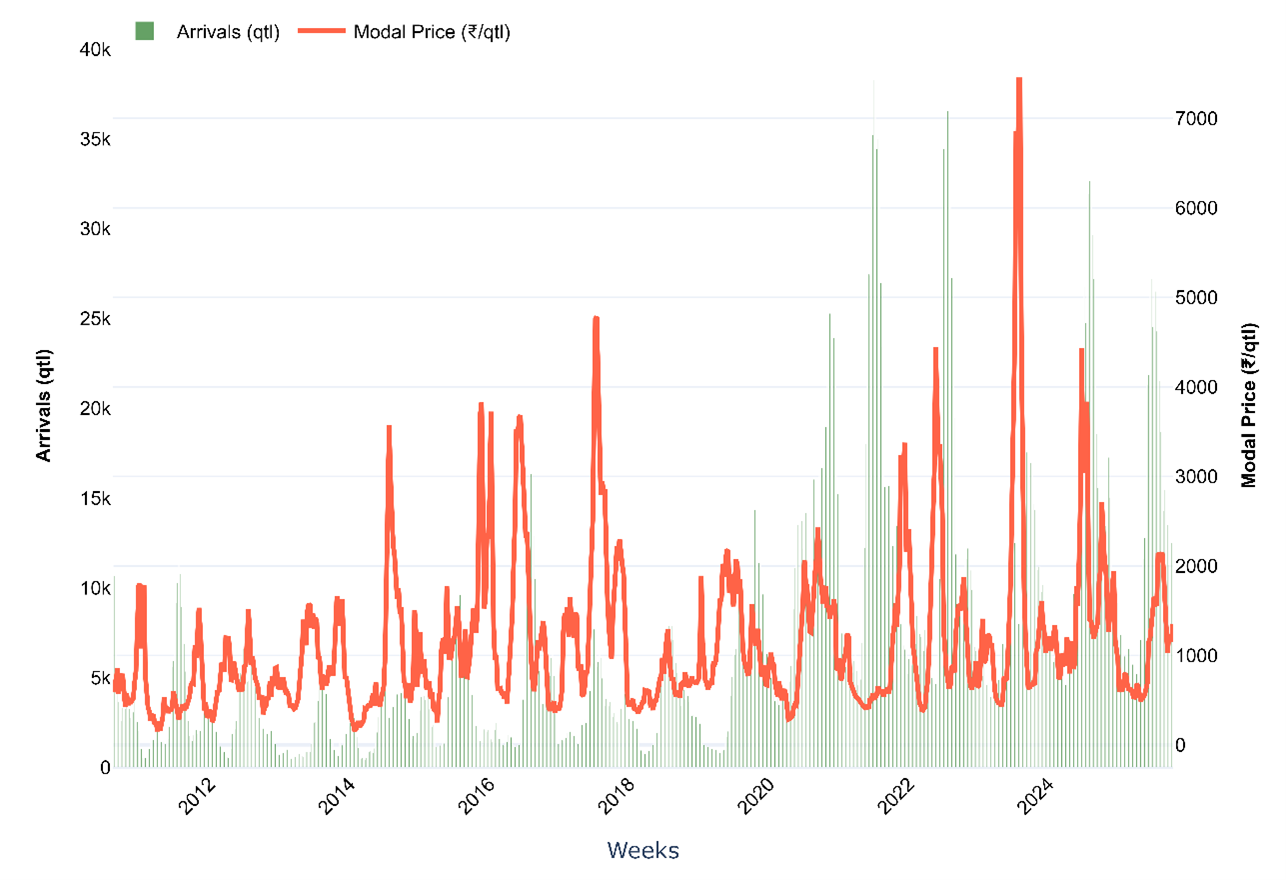}
\caption{Weekly trends of tomato arrivals and modal prices in Kolar market
         (2010--2025).}
\label{fig:weekly_trends}
\end{figure}

Figure~\ref{fig:weekly_trends} shows weekly arrivals and modal prices
(2010--2025), revealing the expected inverse relationship: large influxes were
usually followed by price troughs (weeks 18--22 and 40--48). Deviations from
this pattern --- high arrivals coupled with high prices or low arrivals with
low prices --- were common, underscoring that factors such as quality
heterogeneity, festival demand, weather shocks, or inter-market arbitrage also
influenced prices \cite{acharya2012,miljkovic2024}. The coexistence of multiple
arrival-price combinations confirmed the non-linear nature of price formation
in perishable produce markets, as documented for Indian tomatoes by
\cite{ali2023}.

\begin{figure}[htbp]
\centering
\includegraphics[width=0.95\linewidth]{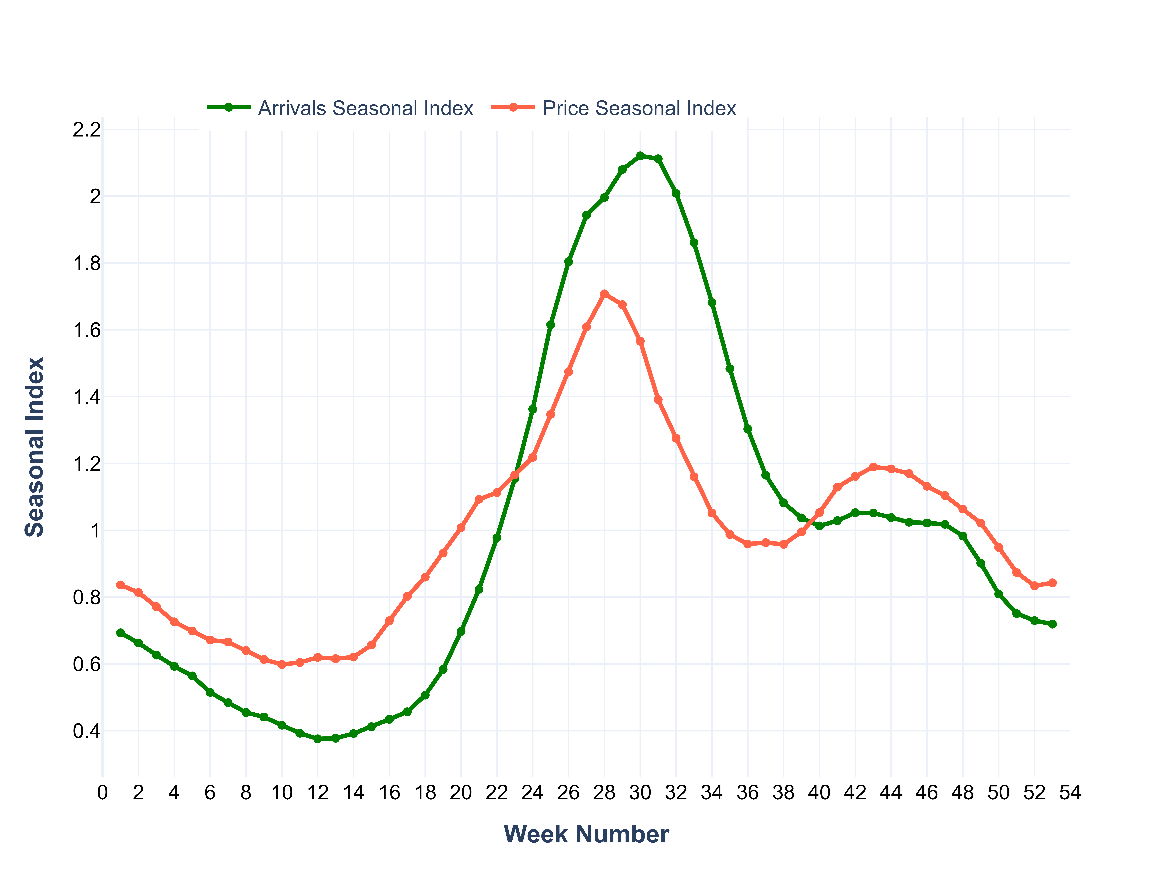}
\caption{Weekly seasonal indices of tomato arrivals and modal prices in Kolar
         market (2010--2025).}
\label{fig:seasonal_index}
\end{figure}

Figure~\ref{fig:seasonal_index} shows weekly seasonal indices (2010--2025),
indicating a bi-modal arrival cycle with an early peak in weeks 5--15 and a
larger one in weeks 40--48, reflecting Kolar's two main harvest windows. Prices
moved inversely, troughing during supply surges and rebounding once arrivals
tapered off, consistent with standard supply-price dynamics \cite{murthy2009}.
A secondary price spike appeared in the post-peak weeks when residual supply
met festival demand.

However, identical index values across years mask calendar drift: weather
shocks, seedling advancement, or policy bans can shift peaks by several weeks
\cite{chand2010}. Static seasonal dummies therefore risk mis-timing forecasts;
the next section uses Dynamic Time Warping to quantify how much the seasonal
calendar slid from year to year.

\subsection{Inter-annual Pattern Stability: Dynamic Time Warping (2021--2024)}

\subsubsection{Modal Prices}

\begin{figure}[htbp]
\centering
\includegraphics[width=0.95\linewidth]{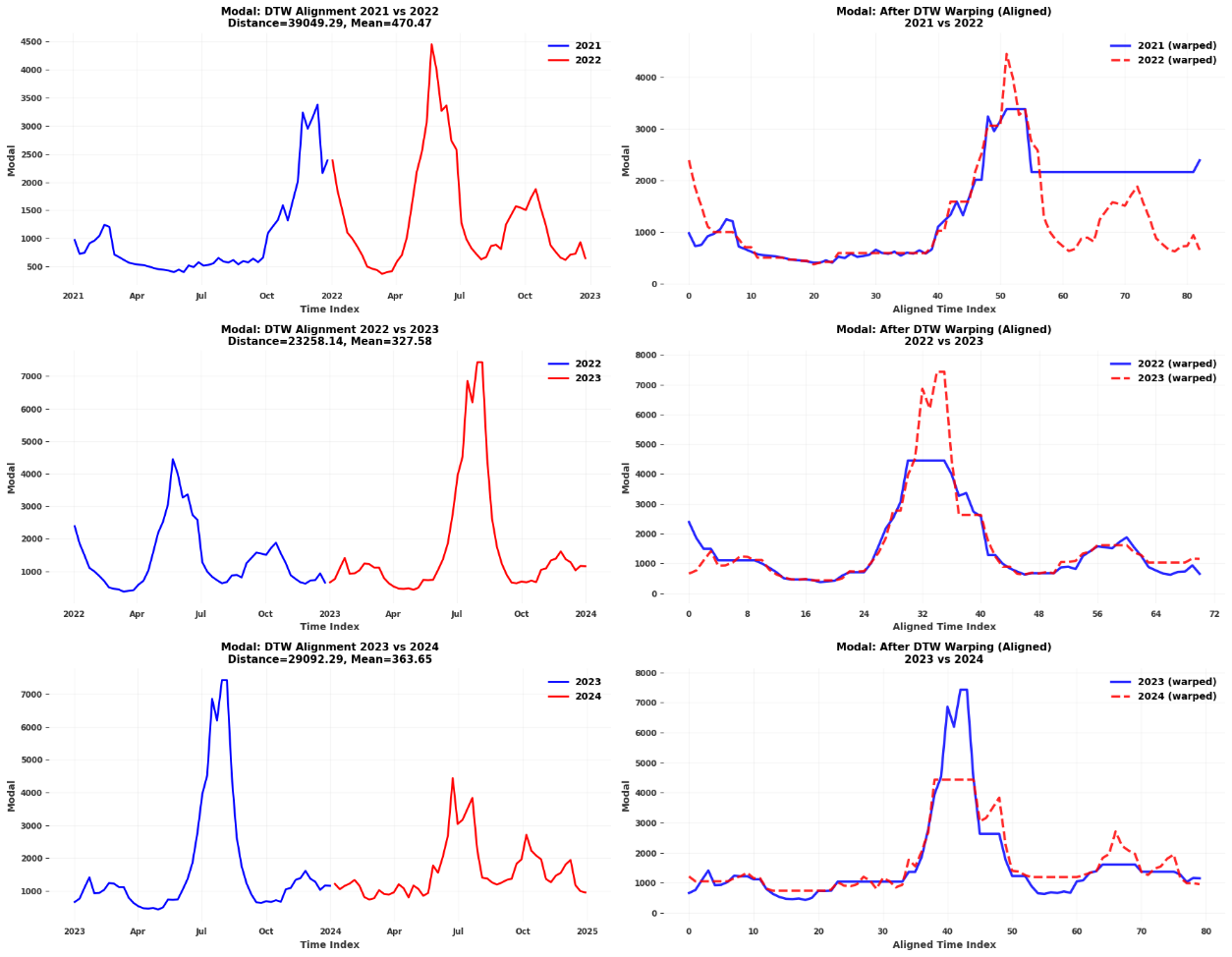}
\caption{DTW analysis of modal price patterns (2021--2024). Left panels: DTW
         alignment paths; right panels: warped (time-aligned) price series.
         Lower DTW distances indicate higher inter-annual pattern similarity.}
\label{fig:dtw_prices}
\end{figure}

DTW analysis of weekly tomato prices from 2021 to 2024 revealed notable
inter-annual differences in pattern similarity. This method was chosen to
capture post-pandemic shifts in market behaviour while maintaining statistical
robustness. Consistent with prior studies demonstrating DTW's ability to detect
structural similarity despite temporal misalignment \cite{berndt1994,sakoe1978},
the analysis successfully aligned seasonal peaks and troughs even when their
timing varied across years (Table~\ref{tab:dtw_prices};
Fig.~\ref{fig:dtw_prices}).

The comparison showed the strongest alignment during 2022--2023 (DTW distance:
23,258; mean distance: 327.58), indicating that despite sharply higher prices
in 2023 (maximum \rupee~7,429 vs.\ \rupee~4,447 in 2022), seasonal patterns
remained consistent. This decoupling of magnitude and timing has been observed
in other volatile agricultural markets \cite{miljkovic2023}. In contrast,
2021--2022 exhibited the weakest similarity (DTW distance: 39,049; mean
distance: 470.47), reflecting a structural break in price behaviour during
post-pandemic recovery, characterised by supply chain disruptions and altered
production cycles.

Path length analysis supported these observations: 2022--2023 required only 71
alignment points versus 83 for 2021--2022, indicating tighter temporal
correspondence \cite{wang2020}. Moderate similarity in 2023--2024 (DTW
distance: 29,092; path length: 80) suggested post-shock stabilisation, with a
higher price floor (\rupee~734 vs.\ \rupee~429) and reduced volatility
\cite{gilbert2010}.

Overall, these results confirm that seasonal timing in tomato prices is
recurrent but not fixed. Variation in DTW distances (23,258--39,049) highlights
the need for forecasting models robust to both magnitude shocks and structural
breaks \cite{makridakis1998,hamilton1989}. Practically, DTW can inform training
data selection, allowing greater weight to years with higher temporal similarity
(e.g., using 2023 data to forecast 2024), and can be operationalised through
DTW-weighted ensemble learning or adaptive neural network training
\cite{yadav2025}.

\begin{table}[htbp]
\centering
\caption{Dynamic Time Warping analysis summary for modal prices (2021--2024).}
\label{tab:dtw_prices}
\adjustbox{max width=\textwidth}{%
\begin{tabular}{lccccccr}
\toprule
\textbf{Year pair} & \textbf{Obs.\ (Y1/Y2)} &
\textbf{Price range Y1 (\rupee)} & \textbf{Price range Y2 (\rupee)} &
\textbf{DTW dist.} & \textbf{Mean dist.} & \textbf{Path len.} & \textbf{Rank} \\
\midrule
2021 vs 2022 & 52/52 & 401--3{,}377  & 371--4{,}447  & 39{,}049 & 470.47 & 83 & 3 \\
2022 vs 2023 & 52/53 & 371--4{,}447  & 429--7{,}429  & 23{,}258 & 327.58 & 71 & 1 \\
2023 vs 2024 & 53/52 & 429--7{,}429  & 734--4{,}434  & 29{,}092 & 363.65 & 80 & 2 \\
\bottomrule
\end{tabular}}
\smallskip\\
\small\textit{DTW dist.\ = total cumulative alignment cost (lower = more
similar); Mean dist.\ = average per-point distance after alignment;
Path len.\ = number of alignment points (indicates degree of time warping).}
\end{table}

\subsubsection{Arrivals}

\begin{figure}[htbp]
\centering
\includegraphics[width=0.95\linewidth]{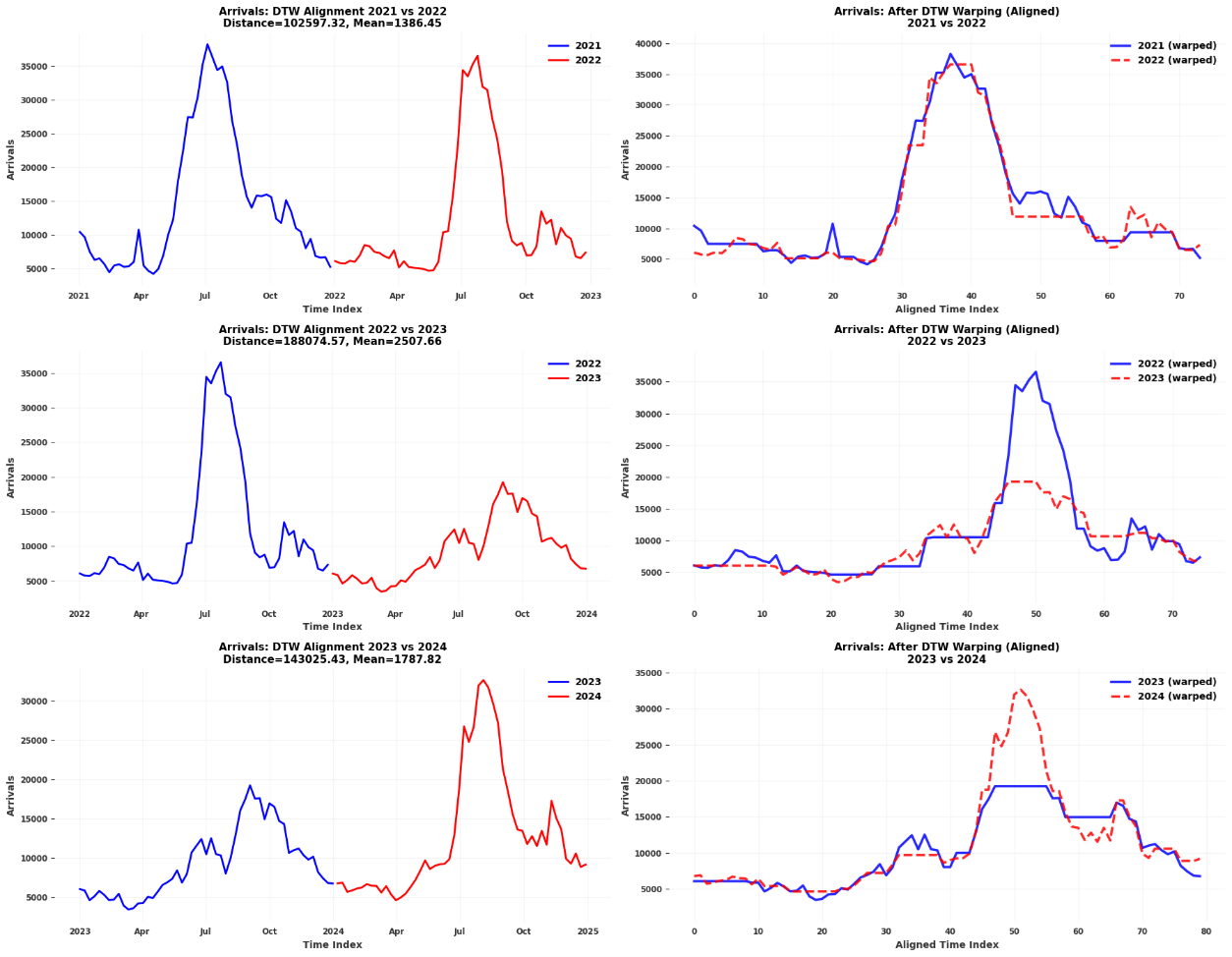}
\caption{DTW analysis of arrivals patterns (2021--2024).}
\label{fig:dtw_arrivals}
\end{figure}

DTW analysis of weekly tomato arrivals (Table~\ref{tab:dtw_arrivals};
Fig.~\ref{fig:dtw_arrivals}) revealed patterns contrasting with price dynamics.
The 2021--2022 comparison showed the highest similarity (DTW distance: 102,597;
mean distance: 1,386.45), indicating that weekly supply timing and magnitude
were highly consistent. This suggests stable production and harvesting
trajectories despite the structural break observed in prices during the same
period. Similar stability in arrivals during post-pandemic recovery has been
reported for other perishable commodities \cite{ali2023}.

In contrast, 2022--2023 exhibited the weakest alignment (DTW distance: 188,074;
mean distance: 2,507.66), reflecting substantial shifts in arrival patterns,
likely due to supply shocks, production decisions influenced by prior-year
prices, or climatic variability. While prices remained temporally aligned in
this period, arrivals diverged, a phenomenon previously attributed to storage
and inventory management strategies where traders adjust release patterns to
stabilise prices \cite{acharya2012}. Inter-regional imports or market channel
substitutions may also have contributed.

The 2023--2024 comparison showed intermediate similarity (DTW distance: 143,025;
path length: 80), suggesting partial recovery of typical supply patterns. Path
length variability (74--80 alignment points) was less pronounced than for
prices, indicating that weekly arrivals are generally more stable than price
trajectories, though both exhibit meaningful inter-annual variability.

\begin{table}[htbp]
\centering
\caption{Dynamic Time Warping analysis summary for weekly arrivals
         (2021--2024).}
\label{tab:dtw_arrivals}
\adjustbox{max width=\textwidth}{%
\begin{tabular}{lccccccr}
\toprule
\textbf{Year pair} & \textbf{Obs.\ (Y1/Y2)} &
\textbf{Arrivals range Y1 (qtl)} & \textbf{Arrivals range Y2 (qtl)} &
\textbf{DTW dist.} & \textbf{Mean dist.} & \textbf{Path len.} & \textbf{Rank} \\
\midrule
2021 vs 2022 & 52/52 & 4{,}171--38{,}277 & 4{,}634--36{,}572 & 102{,}597 & 1{,}386.45 & 74 & 1 \\
2022 vs 2023 & 52/53 & 4{,}634--36{,}572 & 3{,}437--19{,}249 & 188{,}074 & 2{,}507.66 & 75 & 3 \\
2023 vs 2024 & 53/52 & 3{,}437--19{,}249 & 4{,}619--32{,}653 & 143{,}025 & 1{,}787.82 & 80 & 2 \\
\bottomrule
\end{tabular}}
\end{table}

Overall, these results highlight the complex, non-linear relationship between
supply arrivals and price formation. Divergent DTW patterns indicate that price
dynamics were influenced not only by immediate supply but also by storage
behaviour, demand fluctuations, and market expectations \cite{miljkovic2024}.
Forecasting models should therefore incorporate both supply-side indicators and
market behaviour variables to capture the full range of factors shaping price
movements.


\section{Conclusion}\label{sec:conclusion}

Dynamic Time Warping showed that seasonal patterns in Kolar tomato prices and
arrivals recur but rarely line up on the calendar. Price trajectories aligned
best in 2022--23 even while recording the highest absolute volatility, whereas
arrival timings were most stable in 2021--22. This mismatch confirms that price
formation is driven by more than immediate supply; storage behaviour,
inter-regional flows, and market expectations regularly shift the seasonal
calendar. DTW can quantify inter-annual pattern drift and separate calendar
shocks from pure magnitude spikes --- information that seasonal indices alone
cannot provide. Farmers can use early-season DTW alerts to adjust planting or
staggered harvesting; traders and cold-store managers can time inventory
release; policymakers can replace rigid calendar-based interventions with
adaptive triggers that activate when the current year diverges beyond a
DTW-distance threshold.

Limitations include the short four-year window, single-market focus, and
absence of a formal bootstrap significance test or an out-of-sample forecast
evaluation using DTW-weighted training sets. Extending the series and embedding
DTW distances into transformer architectures (PatchTST, Informer) as
patch-weighting or adaptive-window features is an obvious next step. Combining
DTW diagnostics with modern deep-learning forecasts offers a practical route to
more resilient decision-support in volatile perishable value chains.

\bibliographystyle{unsrt}
\bibliography{references}

\end{document}